\def\simlt{$\; \buildrel < \over \sim \;$}
\def\ltsima{\lower.5ex\hbox{\simlt}}
\def\simgt{$\; \buildrel > \over \sim \;$}
   
\def\37{NGC~3783}
\def\55{NGC~5548}
\def\p{$\pm$}

\def\ltsima{$\; \buildrel < \over \sim \;$}
\def\simlt{\lower.5ex\hbox{\ltsima}}
\def\gtsima{$\; \buildrel > \over \sim \;$}
\def\simgt{\lower.5ex\hbox{\gtsima}}

\def\etal{{\it et~al.~}}

\def\flux{\;\rm erg\ cm^{-2}\ s^{-1}}
\def\csq{$\chi^2$}

\documentclass{aa}
\usepackage{graphicx}

\begin{document}

%\thesaurus{ }

\title{The BeppoSAX broad-band spectrum and variability of the Seyfert 1 NGC 3783}

\author{ A. De Rosa \inst{1}
\and L. Piro \inst{1}
\and F. Fiore \inst{2}
\and P. Grandi \inst{1}
\and L. Maraschi \inst{4}
\and G. Matt \inst{3}
\and F. Nicastro \inst{5}
\and P.O. Petrucci \inst{4}
}

\offprints{Alessandra De Rosa: derosa@ias.rm.cnr.it}

\institute{
{Istituto di Astrofisica Spaziale, 
C.N.R.,Via Fosso del Cavaliere, Roma, Italy}
\and
{Osservatorio Astronomico di Roma, Monteporzio Catone, Italy }
\and
{Dipartimento di Fisica, Universit\`a degli Studi ``Roma Tre'', Via della Vasca
Navale 84, I--00146 Roma, Italy}
 \and{Osservatorio Astronomico di Brera, Milano, Italy}
\and
{Harvard-Smithsonian Center of Astrophysics, Cambridge MA 02138 USA}}
\date{Received ; Accepted }

\abstract{BeppoSAX observed the Seyfert 1 galaxy NGC~3783 for $\sim$ 5 days
from 1998 June 6 to 1998 June 10.
The average flux during the observation was 
F$_{2-10 keV}\sim 6\times 10^{-11} \flux$.
The long exposure provided a very high quality spectrum that exhibits  
all the typical features of Seyfert 1s.
The high energy cut-off of the intrinsic continuum, at 
E$_c=340\pm^{560}_{107}$ keV, was detected for the first time in this object.
During the observation the source showed modest flux variations
($\sim$ 20\% in 2-10 keV) and spectral variability that can be completely  
explained by a change of the intrinsic slope $\Delta \Gamma\sim$ 0.1.
NGC~3783 is one of the few Seyfert 1s that exhibits a 
soft emission component in the BeppoSAX spectrum. 
This soft excess is well reproduced by a black body component with 
temperature kT$\sim$0.2 keV.
A warm absorber is also required to fit the soft X-ray spectrum.
Both reprocessing features, the Compton reflection hump and the 
Fe emission line, were detected in our observation.
In addition to a narrow iron line (as observed by Chandra) a broad 
component with $\sigma=0.72\pm^{1.28}_{0.27}$ keV and EW=115$\pm$76 eV, is 
required to fit the line.
The observed Compton hump and the Fe line are well reproduced with the ionized
($\lg\xi=2.73\pm^{0.04}_{0.05}$) disc reflection model of 
Ross \& Fabian.
In this case the soft excess observed below 2 keV can be accounted for by the 
disc emissivity and no additional soft X-rays component is required
to fit the spectrum. 

}

\maketitle

\section{Introduction}

Intrinsic spectral variability has been observed in several
Seyfert 1 galaxies: NGC~5548 (\cite{N00}),  NGC~4151 (Piro \etal 2002), 
IC~4329A (\cite{done2000}), NGC~7469 (Nandra \etal 2000), 
MCG-6-30-15 (Vaughan \& Edelson 2001).
The usual behaviour is a steepening of the primary
power law when the source brightens.
The favoured model for the X-ray emission observed in AGN comprises a 
hot, optically-thin plasma (the corona) which up-scatters to X-ray energies 
the soft photons produced in a cold, optically-thick accretion disc 
(\cite{HM91}).
In addition the optically thick disc reprocesses and re--emits part of the
Comptonized flux producing the characteristic reprocessing features 
(the Compton reflection hump and the K$\alpha$ Fe line).
In a disc+corona emission model the intrinsic variability has to be 
associated with a change of some properties either of the emitting corona 
(temperature kT$_e$, optical depth $\tau$ or geometry, see \cite{HMG97}) 
or the accretion disc (flares) with the X-ray flux.
A detailed Comptonization model applied to the BeppoSAX long look of NGC 5548
(Petrucci \etal 2000) showed that the steepening of the spectrum during a 
flare in the central part of the observation was probably associated with 
an increase of the soft photon luminosity, the hard photon luminosity 
remaining constant. 

The presence of a broad iron line in the 
X-ray spectra of Seyfert galaxies (Fabian \etal 2000) is suggested by        
ASCA (\cite{nandra97}, \cite{tanaka95}, \cite{yaq2002}), BeppoSAX (Guainazzi \etal 1999) and
XMM (Wilms \etal 2001) observations.
Nevertheless a narrow iron line component is clearly detected in 
the Chandra (\cite{kaspi2}, Yaqoob \etal 2001) and XMM observations
(Reeves \etal 2001, Pounds \etal 2001). 
This narrow component could be attributed either to an optically thin 
medium (like the BLR), and in this case no Compton reflection should be 
associated to it, or to an optically thick medium with N$_H>10^{24}$cm$^{-2}$ 
(as the molecular torus like that observed in the Sy2 s, \cite{matt01a}), 
which should contribute to the observed Compton reflection 
(\cite{ghisellini94}). 
At least in the case of NGC~4051 (Guainazzi \etal 1998)
it is possible to associate the narrow iron line component to a distant 
Compton-thick matter.

The {\it Broad band spectral variability in bright Seyfert 1 galaxies} 
program, carried out by BeppoSAX in the last 3 years, 
exploits the unique capability of 
BeppoSAX to provide spectral informations up to 200 keV. 
The primary goals of this program are to investigate the correlation 
of the intrinsic parameters ($\Gamma$ and $E_c$) with the luminosity and 
to characterize the diverse components of Seyfert 1 spectra
from 0.1 to 200 keV, by studying their response to changes of the 
intrinsic luminosity.

The bright Seyfert 1 galaxy \37 was selected as a target in the cited 
BeppoSAX program because: 
a) it is bright enough that the spectrum up to 
about 100 keV ($F(2-10 keV)>2\times 10^{-11}$ erg cm$^{-2}$ s$^{-1}$
could be measured;
b) it is characterized by variations of a factor of two or more on the time 
scale of days. This time scale is appropriate to check possible 
correlations in the variations of the spectral components.
The extensively studied X-ray spectrum of this object is characterized by 
narrow and broad features in addition to the intrinsic continuum.
Deep Oxygen absorption edges (between 0.5 and 1 keV) were detected by ROSAT 
(\cite{turneretal93}) and ASCA 
(\cite{G98a}, \cite{G98b}) which also observed a variable warm absorbing gas 
along the line of sight. An OVII emission line, likely originating in 
the same warm gas, was also detected in the first ASCA 
observation (in 1993). 
The Chandra high energy resolution 
observation (\cite{K00}) identified, in addition to many narrow 
absorption lines, several weak emission lines mainly from O and Ne, 
suggesting the presence of two absorption components (outflowing by the 
central source), that differ by an order of magnitude in their ionization 
parameters.
A Compton reflection hump and a Fe K$\alpha$ emission line were
detected by GINGA (\cite{np94}). 
A Chandra observation (\cite{kaspi2}) detected a
narrow and unresolved iron line, with an upper limit to the FWHM of
3250 km s$^{-1}$. These data could not constrain any broad iron line, though 
they are consistent with the broad line found in previous ASCA observations.
\newline

In this paper we present the analysis of a BeppoSAX observation of \37.
In section \ref{observation} we describe the observation and the data 
reduction.
In section \ref{spec tot} we report the spectral analysis of the data 
integrated over the total observation while in section \ref{spec var}
the spectral variability analysis is presented.
The results are discussed in section \ref{discussion}. 
A summary is given in section \ref{summary}.

\section{Observations and data analysis}
\label{observation}

\37 was observed by BeppoSAX (\cite{boella97}, \cite{piro95}) for $\sim$ 
5 days ($t_{exp}\sim 154$ ks) 
from 1998 June 6 to 1998 June 10.
The LECS, MECS and PDS data reduction followed the standard procedure
(\cite{cook}).
The PDS spectra were filtered with fixed rise time.
We extracted the spectrum within a circular region centered on the 
source with a radius of 4' and 6' for MECS and LECS respectively.
The background was extracted from event files of source-free regions 
(``blank fields''). Spectral models 
were fitted to the data using the XSPEC package. All quoted 
uncertainties correspond to 90\% confidence interval for one interesting 
parameter ($\Delta$\csq=2.71).
The models included a normalization factor for each instrument
to take into account possible miscalibrations. 
We allowed the PDS/MECS normalization to vary between 
0.77 and 0.95, while the LECS/MECS normalization ratio was
running between 0.7 and 1 (\cite{cook}).
The journal of the BeppoSAX observation of \37 is shown in Table 
\ref{journal}.

\begin{figure*}
\centering
\includegraphics[height=11.0cm,width=8.5cm]{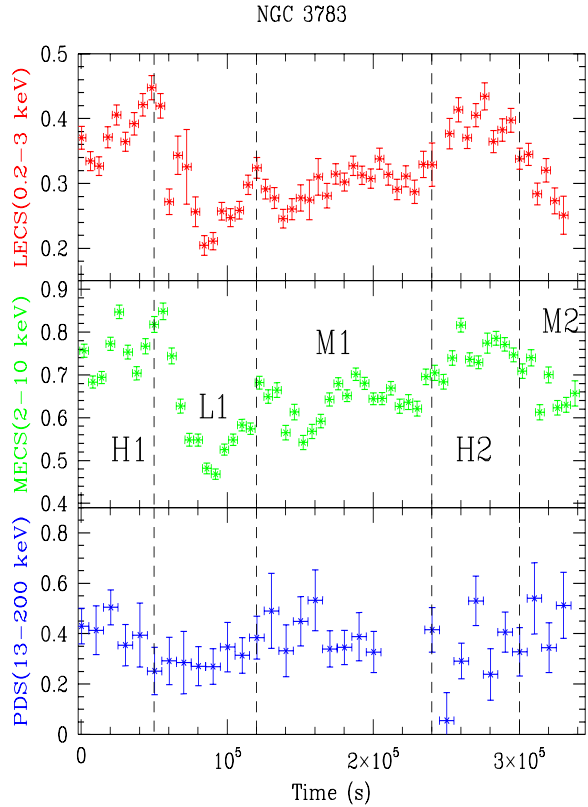}
\caption[]{LECS, MECS and PDS lightcurves of BeppoSAX observation 
of \37. The selected different flux level states are also indicated.}
\label{light}
\end{figure*}

\begin{table*}
\caption{Journal of the BeppoSAX observation of NGC~3783.  
The different flux level states defined in Fig. \ref{light} are 
indicated in each row. The spectral variability is discussed in Sect. 
\ref{spec var}.} 
\begin{flushleft}
\begin{tabular}{l c c c c c c c}
\noalign{\hrule}
\noalign{\medskip}
\hline
& & & & & & & \\
State & \bf Exposure (ks) & \bf $^\diamond$Count Rate (cts/s) & \bf $^\dagger$S/N & $^\bullet$C &  $^\star{\chi}^{2}$(dof) & $^\ast$P$_{{\chi}^{2}}$ \\
& & & & & & & \\
\hline
\bf TOT  & 153870 & 0.758 \p 0.002 & 341 & & & & \\
\hline
\bf L1   & 27403 & 0.586 \p 0.005 & 127 & & & & \\
%& & & &  & & & \\
\bf H1   & 35630 & 0.823 \p 0.005 & 171  & & & & \\
%& & & &  & & & \\
\bf M1  & 64257 & 0.709 \p 0.003 & 213  & & & &  \\
%& & & & &  & & \\
\bf M2   & 13311 & 0.720 \p 0.007 & 99 & & & &  \\
%& & & & & & & \\
\bf H2   & 26791 & 0.851 \p 0.006 & 151  & & & & \\
%& & & &  &\\
\bf H = H1+H2 & 62421 & 0.835 \p 0.004 & 228 & 1.02 & 89.63(82) & 0.26 \\
%& & & &  & & \\
%\bf L1M2=L1+M2& 40714 & 0.633 \p 0.004 & 161 & 1.4 & & \\
%& & & &  & & & \\
\bf L=M1+M2 & 77568 & 0.713 \p 0.003 & 235 & 1.07 & 109.5(82) & 0.02 \\
%& & & &  &\\
\bf ratio H/L & & & & 1.254 & 222.3(82) & $<<10^{-3}$ \\
\hline
\end{tabular}
\end{flushleft}
\small{$^\diamond$ In 1.5-10 $keV$.}
\newline
\small{$^\dagger$ Signal to noise ratio in 1.5-10 $keV$.}
\newline
\small{$^\bullet$ best fit value when the ratio of the two spectra is modeled 
with a constant} 
\newline
\small{$^\star$ $\chi^2$ value for a constant model with const=C}
\newline
\small{$^\ast$ Probability of exceeding $\chi^2$. }
\label{journal}
\end{table*}
In Figure \ref{light} we show the LECS (0.2-3 keV), 
MECS (2-10 keV) and PDS (13-200 keV) light curves.
The source showed rapid variability both in the soft (0.2-3 keV) 
and medium (2-10 keV) energy ranges, with the larger amplitude at 
low energies.
The 2-10 keV and 0.2-3 keV flux variations were up to 
a factor 1.6 and 2, respectively, on a time scale of about half a day.

%%%%%%%%%%%%%%% lightcurves and ratio %%%%%%%%%%%%%%%%%%%%%%

\section{The broad band spectral analysis: 0.1-200 keV}
\label{spec tot}
\begin{figure*}
\centering
\includegraphics[height=11.0cm,width=8.5cm,angle=-90]{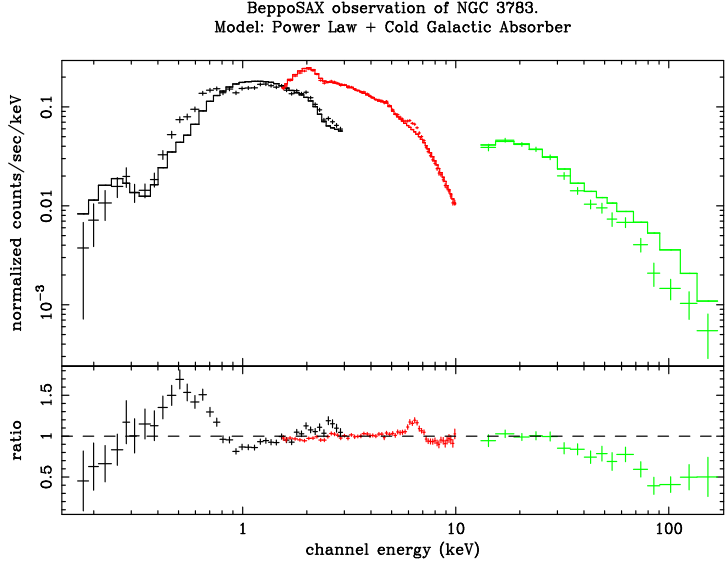}
\caption[]{LECS, MECS and PDS data (upper panel), and data/model ratio 
(lower panel) in the case of an emission continuum fitted with a
power law absorbed by a cold galactic gas (model A).}
\label{spo}
\end{figure*}

In this section we present the spectral analysis of the 
data integrated over the total observation of \37 (t$_{exp} \sim$ 154 ks).
We fitted the LECS, MECS and PDS data in the energy range 0.15-3 keV, 
1.5-10 keV and 13-200 keV respectively.
In Figure \ref{spo} we show the 
LECS, MECS and PDS spectra and the data/model ratio after 
fitting with (model A) 
a power law absorbed by the Galactic hydrogen column density 
($N_H^{gal}=9.6\times 10^{20}$ cm$^{-2}$, \cite{murphy96}).  
The fit is very bad ($\chi^2$/dof=818.6/114, with $\Gamma$=1.58\p0.06).
Several deviations of the data from the model are apparent:
an excess of counts at $\sim$ 0.6 keV, 
a deficit below 2 keV, an emission line 
at $\sim$ 6.4 keV and another deficit above 100 keV.

\begin{figure*}
\centering
\includegraphics[height=11.0cm,width=8.5cm]{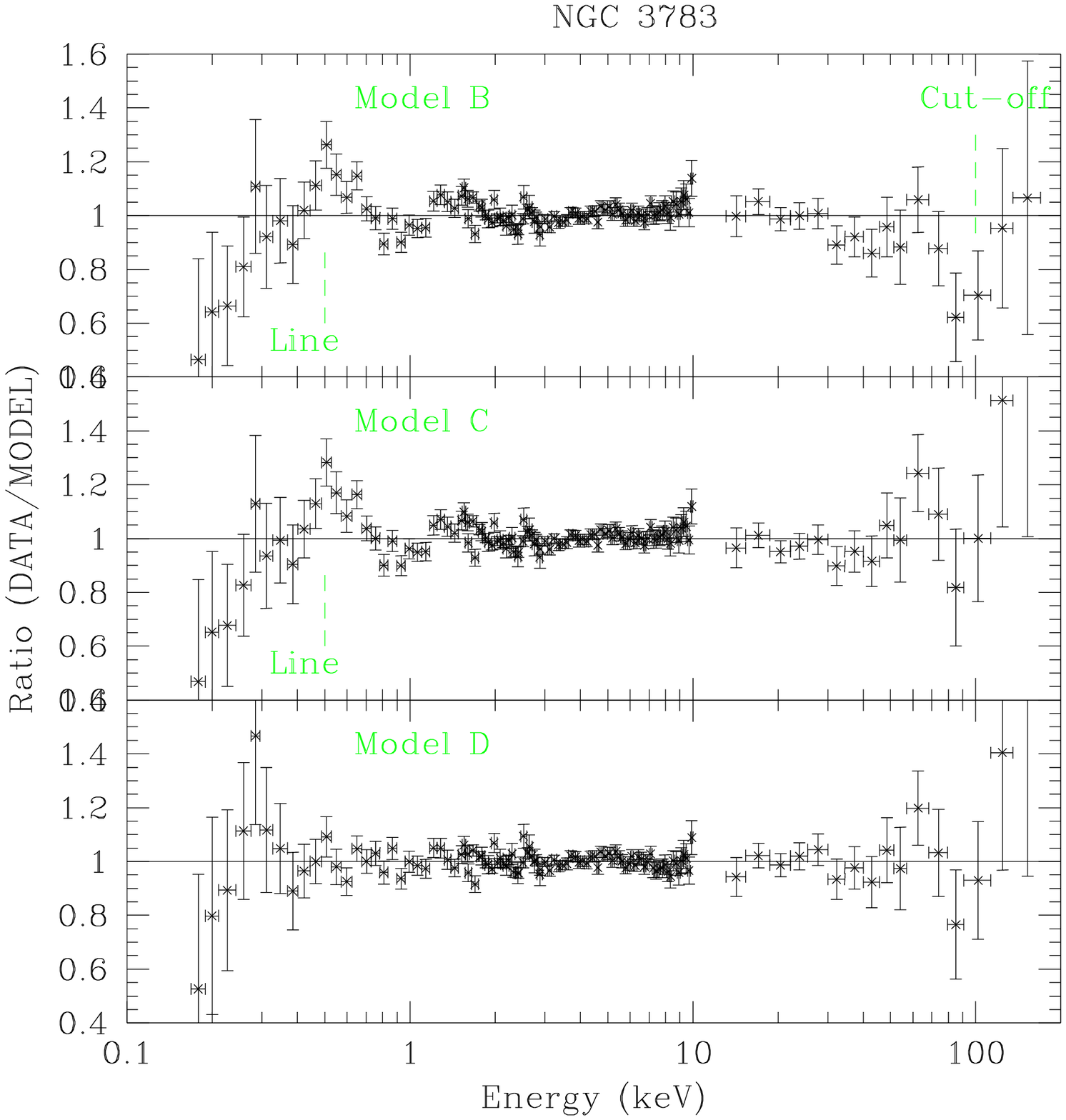}
\includegraphics[height=11.0cm,width=8.5cm]{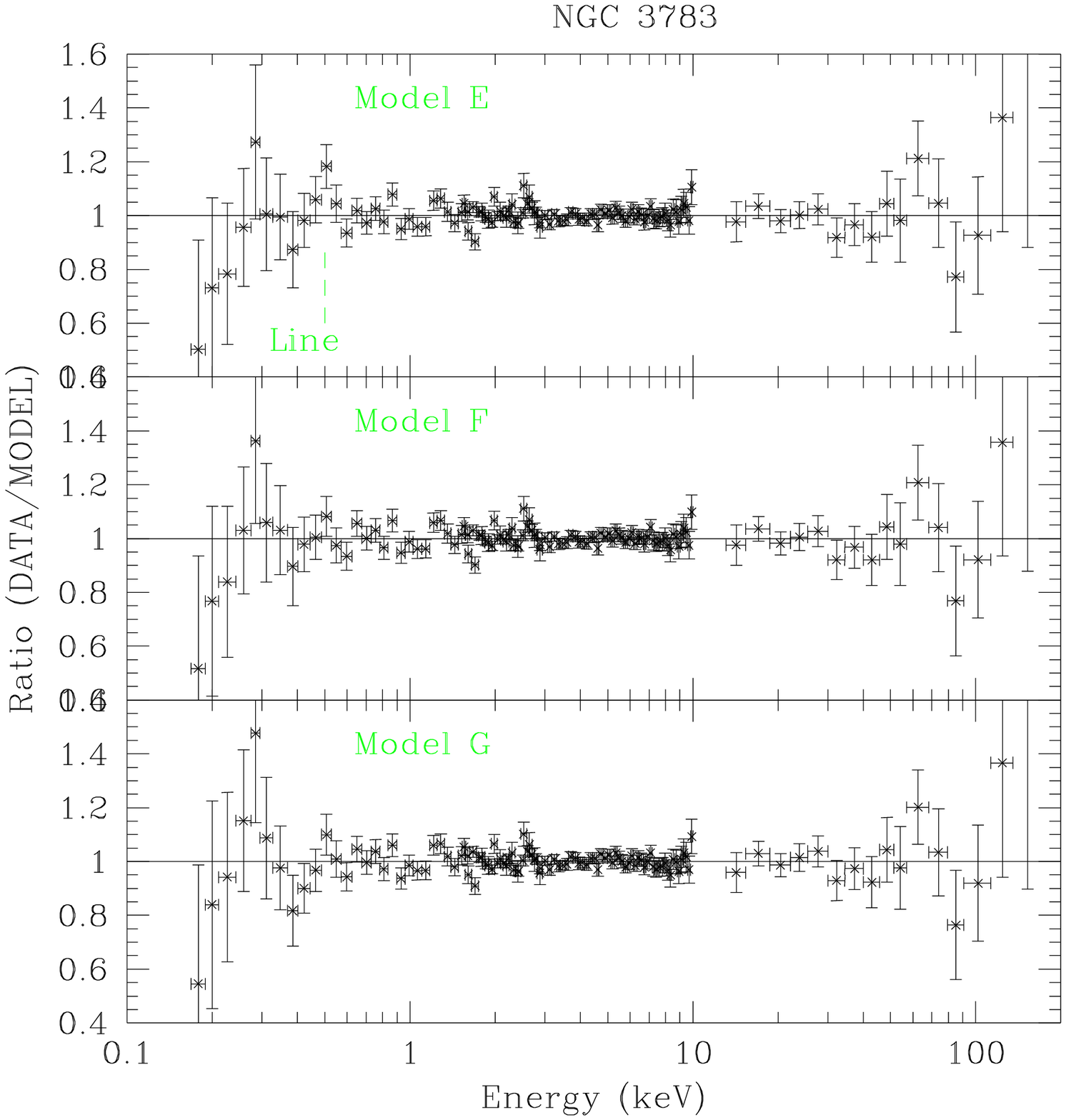}
\caption[]{Ratio data model for the different spectral emission models
employed to fit the \37 BeppoSAX data. See Table \ref{fit tot2}.}
\label{ratio}
\end{figure*}

We then fitted the data with a model (B) including the known spectral 
components already detected in \37: a warm absorber (\cite{turneretal93}), 
a Compton reflection hump and an iron emission line (modeled with a
gaussian) at $\sim$ 6.4 keV (\cite{np94}).
The warm gas, assumed to be in photoionization equilibrium, is characterized  
by two parameters: the ionization parameter $U$, defined as the ratio of the
number density of ionizing photons to hydrogen nuclei, and 
the hydrogen column density N$_H$. The description and discussion of the 
photoionization code we employed will be given in Sect. \ref{se}.
To fit the Compton hump we use here a cold disc reflection model 
(PEXRAV in XSPEC, \cite{MZ95}). The reprocessed spectrum is characterized by 
the reflection fraction $R=\Delta\Omega/2\pi$ i.e. the solid angle subtended 
by the reflecting medium. The inclination angle between the line of sight 
and the reflector is assumed i=30$^\circ$.
For the moment neither ionization nor smearing effects 
(gravitationals and kinematic) are taken into account to fit the
reprocessed features. 
We will discuss these cases in Sect. \ref{refl disc}. 
All the line parameters (energy, intrinsic width and intensity) 
are free to vary. The effects of including a narrow 
line component - as observed by Chandra 
(\cite {kaspi2}) - will be discussed in Sect. \ref{refl disc}.   
The result of the fit with the model B is still unacceptable 
($\chi^2$/dof=189.6/108. See Table \ref{fit tot2}).
The ratio data/model B is shown in the upper plot of the left panel
of Figure \ref{ratio}.
This plot shows still some deviations of the model from the data: 
a deficit of counts above 100 keV and an excess at 
energy less than 1 keV.
We therefore added to the primary power law in the model B an exponential 
cut-off (Model C).
The statistical significance of this new spectral component is larger
than 99\% (following a F-test probability for the addition of 
1 interesting parameter we find F=8.14, see Table \ref{fit tot2}).

The excess at low energies may correspond to OVII emission line. 
As suggested by the ASCA observation (\cite{G98a}), 
we fitted this excess with a gaussian emission line.
The model is indicated with D and the best fit parameters with 
the statistical significance of the new spectral component is shown 
in the second column in Table \ref{fit tot2}. 
The ratio data/model D is shown in the lower plot of the left panel in 
Figure \ref{ratio}. 
%
%============== TABELLA SPECTRAL MODEL=======

\begin{table*}
\caption{NGC 3783: Spectral emission models}
\begin{flushleft}
\begin{tabular}{||c|c|lllll||} \hline
~ & ~ & & & & &\\
~ & & \bf Moel B & \bf Model C & \bf Model D & \bf Model E & \bf Model F \\
~ & ~ & & & & & \\ \hline
~ & ~ & & & & & \\
{\bf Warm} & $\lg$(U) & 1.54$\pm^{0.01}_{0.06}$ & 1.60$\pm^{0.01}_{0.06}$ & 1.29$\pm^{0.15}_{0.18}$ & 1.29$\pm^{0.09}_{0}$ & $1.33\pm 0.10$ \\ \cline{2-7}

~  & ~ & & & & & \\ 

{\bf Absorber} & $\lg (N_H)$ & 22.32$\pm^{0.03}_{0.04}$ & 22.31$\pm^{0.03}_{0.06}$ & 21.85$\pm^{0.16}_{0.14}$  & 22.28$\pm^{0.06}_{0.02}$ & $22.26\pm^{0.06}_ {0.08}$\\ \hline

~  & ~ & & & & & \\

{\bf Intrinsic} & E$_c$ (keV) & - & 299$\pm^{307}_{108}$ & 143$\pm^{75}_{36}$ & 307$\pm^{110}_{312}$ & 340$\pm^{560}_{107}$\\ \cline{2-7}

~ & & & ~ & & & \\
{\bf Continuum} & $\Gamma$ & 1.90\p0.04 & 1.89$\pm^{0.03}_{0.03}$ & 1.69$\pm^{0.06}_{0.05}$ &  1.86\p 0.03 & 1.87$\pm^{0.06}_{0.05}$ \\ 
\cline{2-7}

~  & ~ & & & & & \\

~ & $^\diamond$Norm. & 2.2\p 0.1 & 2.1\p 0.1 & 1.6\p 0.1 & 2.1\p 0.1 & 2.1\p 0.1 \\ 
\hline
~&   & ~ & & & &\\

{\bf Reflection} & $^\star$R & 0.75$\pm^{0.20}_{0.17}$ & 0.95$\pm^{0.25}_{0.18}$ & 0.31$\pm^{0.27}_{0.16}$  & 0.71$\pm^{0.20}_{0.28}$ & 0.70$\pm^{0.55}_{0.40}$ \\ \cline{2-7}

~ & ~ & & & & &\\

{\bf and} & E$_L$ (keV) &  6.41 \p 0.04 & 6.41 \p 0.04 & 6.40 \p 0.04 &  $6.39 \pm 0.09$ & $6.39\pm 0.09$ \\ \cline{2-7}

 & & & & & &\\

{\bf Iron Line} & $\sigma_L$ (keV)& 0.53$\pm^{0.21}_{0.11}$ & 0.49$\pm^{0.12}_{0.26}$ & 0.39$\pm^{0.05}_{0.10}$ &  0.38$\pm^{0.11}_{0.09}$ & $0.36\pm^{0.09}_{0.08}$ \\ \cline{2-7}

& & & & & &\\

~& EW$_L$ (eV)& 300 \p 45 & 270 \p 60 & 216 \p 50 & 210\p 45 & 190\p 47\\ \hline
~  & ~ & & & & & \\ 
{\bf OVII} & E$_O$ (keV) & - & - &0.51 \p 0.02  & - &0.52\p0.06 \\ \cline{2-7}
 ~ &  ~ & & & & &\\
{\bf Line} & $\sigma_O$ (keV) & - & - & 0.13$\pm^{0.05}_{0.03}$ & - & $<$ 0.2 \\ \cline{2-7}
~ & ~ & & & & &\\
~ & EW$_O$ (eV) & - & - &240 \p 27 & - & 10 (frozen)  \\ \hline
 ~ &~ & & & & &\\
{\bf Soft} & kT (keV) & - & - & - & 0.21 \p 0.01 &0.22\p0.03 \\ \cline{2-7}
 ~ &~ & & & & &\\
{\bf Emission} & $^{\dagger\dagger}$Norm. &- & - & - &  $4.3\pm^{0.9}_{0.8}$ & 
$3.4\pm^{1.0}_{0.3}$  \\ \hline
 ~ &~ & & & & & \\
& $\chi^2$/dof & 189.6/108 & 176.2/107 & 111.2/104  & 104.7/105 & 101.7/103\\ \hline
 ~ &~ & & & & & \\
& $^\dagger$F value & 59.7 (99.9\%) & 8.14 (99.5\%) & 20.3(99.9\%) & 35.9(99.99\%) & 1.5(77\%)\\
\noalign{\hrule}
\noalign{\medskip}
\end{tabular}
\newline
\small{$^\diamond$ In 10$^{-2}$ photons keV$^{-1}$ cm$^{-2}$ s$^{-1}$,
a 1 keV.} 
\newline
\small{$^\star$ $R=\Delta\Omega/2\pi$ is the $Relative$ $reflection$,
the solid angle subtended by the reflecting medium. The inclination 
angle between the line of sight and the reflector is assumed 
i=30$^\circ$.}
\newline
\small{$^{\dagger\dagger}$ In 10$^{-4}L_{39}/D_{10}^2$, where L$_{39}$ is 
the source luminosity in units of $10^{39}$ ergs/sec and D$_{10}$ is 
the distance to the source in units of 10 kpc.}
\newline
\small{$^\dagger$ In respect to model A (see text) for model B, 
in respect to model B for model C, in respect to model C for model D, 
in respect to model C for model E, in respect to model E for the model F.}
\end{flushleft}
\label{fit tot2}
\end{table*}
%========================================
%
Despite the good $\chi^2$, there is a problem with model D regarding
the origin of the emission line at low energy.
We measure EW$_{oxygen}$=240\p 27 eV (see Table \ref{fit tot2}).
Can this spectral feature be produced in the same gas which produces 
the O edges at E $<$ 1 keV (the warm absorber)?
We suppose the warm absorber to be ionized by the central source and 
in photoionization equilibrium. 
The value of the equivalent width for the OVII emission line 
produced in a gas characterized by  $\lg(U)=1.29\pm^{0.15}_{0.18}$ 
and $\lg(N_H)=21.85\pm^{0.16}_{0.14}$ (as that observed by BeppoSAX),
should not exceed $\sim$ 10 eV (\cite{fab99}).
If collisional ionization is also important, the emissivity from the 
gas strongly increases (\cite{fab99}). In this case the blend of 
the OVII K$\alpha$ triplet is more prominent and the predicted EW 
would increase up to 70 eV, still lower than measured.

The HETG on board Chandra allowed to resolve the OVII
triplet in \37 (\cite{K00}); the total equivalent width 
was $\sim$ 10 eV (even if poorly constrained)
i.e. much smaller than that measured by BeppoSAX.
 
All these considerations suggest that most of the low energy
excess, which modeled with an emission line, was instead 
due to a soft X-ray continuum.
We therefore fitted the excess at low energies with a black body 
component (model E).
The best fit parameters are shown in Table \ref{fit tot2}.
Comparing model E with model C (in which no oxygen line 
was included), we obtain that the black body 
component is required at the 99.99\% confidence level (F=35.9).

\begin{figure*}
\centering
\includegraphics[height=11.0cm,width=8.5cm,angle=-90]{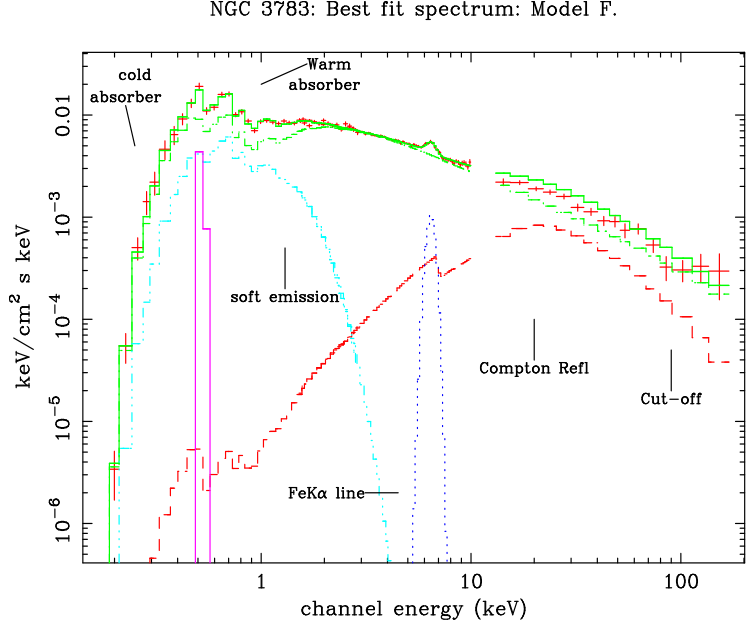}
\caption[]{Unfolded spectrum and the best fit model F (See Table 
\ref{fit tot2}. The contributions to the model of the various additive 
components are also plotted}
\label{spectrum}
\end{figure*}

\begin{figure*}
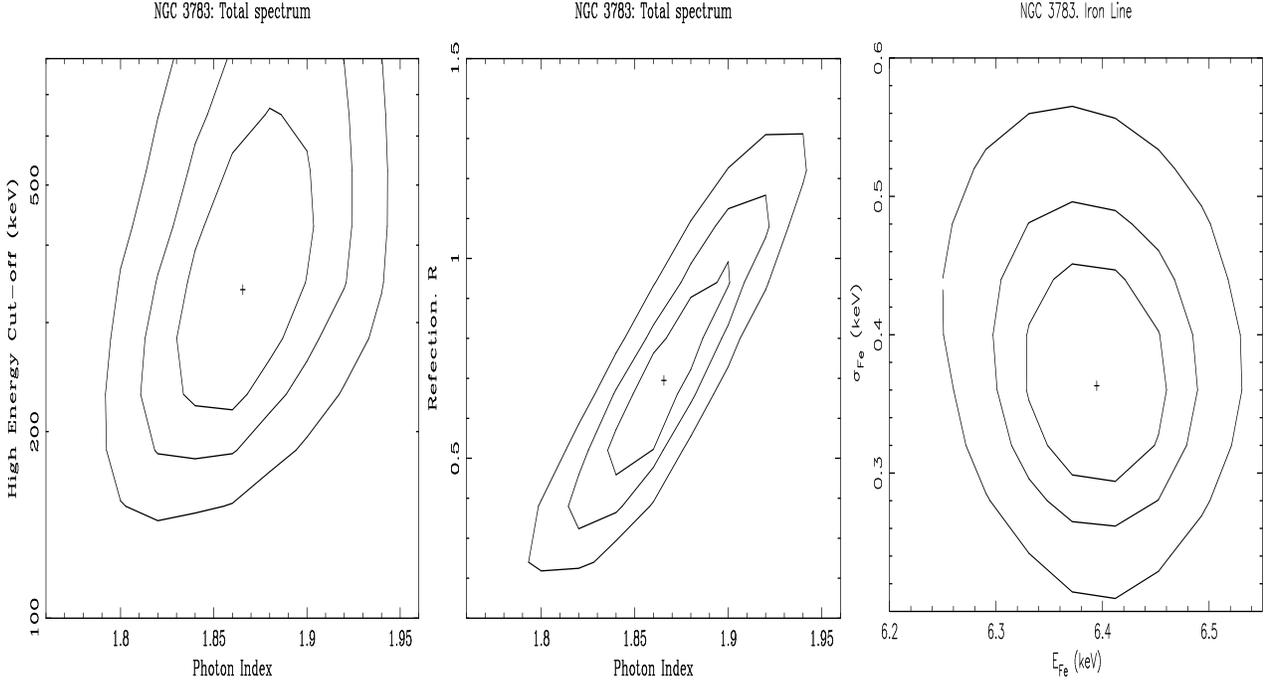

\centering
\includegraphics[height=5.5cm, width=9cm,angle=-90]{ec_gamma_last.ps}
\includegraphics[height=5.5cm, width=9cm,angle=-90]{r_gamma_last.ps}
\includegraphics[height=5.5cm, width=9cm,angle=-90]{sigma_efe_last.ps}
\caption[]{1$\sigma$, 2$\sigma$ and 3$\sigma$ confidence levels E$_c$ vs 
$\Gamma$ (left panel), R vs $\Gamma$ (middle panel) and $\sigma_{Fe}$ 
vs E$_{Fe}$ (right panel) in the total spectrum.}
\label{cont}
\end{figure*}

In the upper plot of the right panel of Figure \ref{ratio} we show the ratio 
data/model E.
When we add a gaussian emission line to the model E (model F), with the 
intensity fixed at the Chandra value, we find that this component
is required at only the 77\% confidence level.
A good fit ($\chi^2/dof=105.1/102$) is also obtained fitting the excess at 
low energies with a thermal bremsstrahlung (the ratio data/model 
for a thermal bremsstrahlung is shown in the last panel on the 
right in figure \ref{ratio}, model G), with temperature kT=0.46\p0.06 keV.
However, in this model the value of the relative reflection is smaller
($R$=0.34 \p 0.23) than expected from the iron line EW$_{Fe}$ ($\sim$ 200 eV).
We therefore conclude that the BeppoSAX spectrum is best fitted by model F.
The best fit model F is shown in Figure \ref{spectrum} where each spectral 
component is labelled.
We plot in Figure \ref{cont} the intervals of confidence for the parameter  
E$_C$ and $\Gamma$ (left panel), $R$ and $\Gamma$ (middle panel)
and $\sigma_{Fe}$ and $E_{Fe}$ (right panel)

\begin{figure*}
\centering
\includegraphics[height=8.5cm,width=8.5cm]{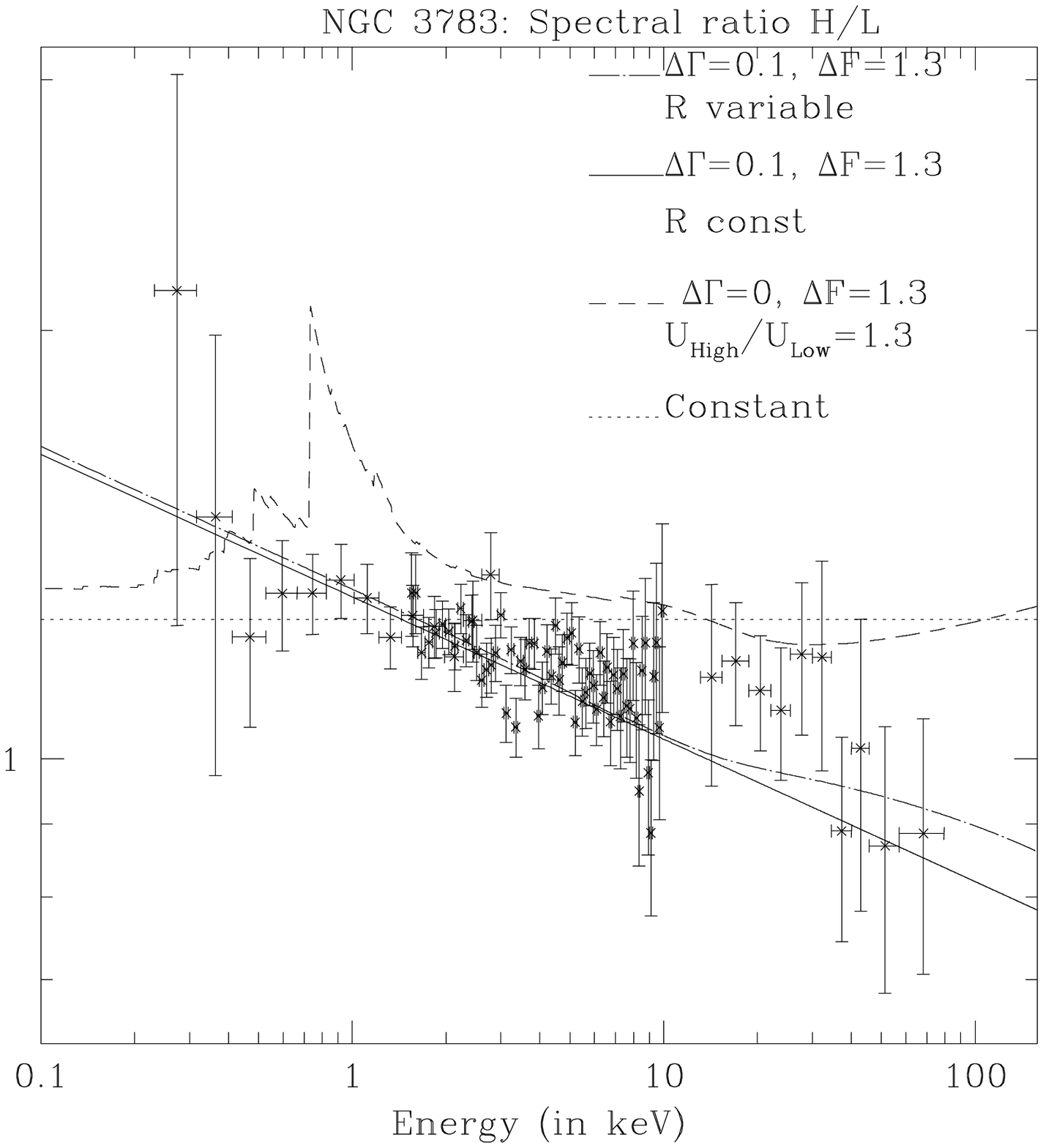}
\caption[]{Ratio of spectra extracted by different flux level states. 
The plotted curves represent the expected behaviour of the ratio 
when the variations are produced by change of the intrinsic continuum, either
with relative reflection R=const (solid line) or with 
relative reflection R variable (dot-dash curve), 
and by a change of the warm gas (dash curve).
The dot line is the best fit value when the ratio is fitted with a constant. 
See text for detail.}
\label{spec ratios}
\end{figure*}

\section{Spectral variability analysis}
\label{spec var}

\subsection {A model independent variability analysis}

To investigate spectral variability,
we extracted spectra from different flux level states of 
the source during the BeppoSAX observation (see Table \ref{journal}, 
see Figure \ref{light}).

The states we selected were: High 1 (H1), Low 1 (L1), 
Medium 1 (M1), High 2 (H2) and Medium 2 (M2). 
To increase the statistics and to get comparable S/N 
ratios of the spectra extracted form the different flux levels, 
we added together those with similar flux. 
We fitted the ratios of the various spectra with a constant model, $r(E)=C$, 
to search for spectral variability. 
The best fit values for C are shown in Table 
\ref{journal}. The ratios H1/H2 and M1/M2 are both consistent with a 
constant value C$\sim$1 (see Table \ref{journal}). 
Then we create a new high state H=H1+H2, and a new low state L=M1+M2.
In Figure \ref{spec ratios} we present the ratio between
H and L spectra. We can reject the hypothesis of constant value for 
this ratio at $>3\sigma$ confidence level (see Table \ref{journal}). 
The plotted curves represent the expected spectral ratio in different cases.
The variations can not be accounted for by a change of the warm absorber 
status (e.g. ionization parameter U). 
In fact we detected deviations of the spectral ratio from the constant model
also above 2 keV (see Figure \ref{spec ratios}), where the warm gas 
is transparent to the radiation and no features can be produced by this 
material.
In Figure \ref{spec ratios} we also plotted the expected ratio if the only 
parameter which changes in the different flux states is the ionization of 
the warm gas (dash line). 
We suppose that the ionization structure is dominated by the intrinsic 
flux between 0.5 and 2 keV, this because the opacity of the gas is dominated 
by K shell absorption edges due to O VII and O VIII (the two most 
abundant ions of the most abundant element). So we rescaled U assuming 
$U_{H}=U_{L}(L_{0.5-2 keV}^{H}/L_{0.5-2 keV}^{L}$).
In this case the absolute normalization of the 
Compton reflection is kept constant in the H and L states.
The other curves in Figure \ref{spec ratios} represent the expected ratio 
if the observed variations were produced by:

1) a change of the intrinsic slope of $\Delta \Gamma=0.1$ (dot - dash curve).
The intensity of the reflected radiation is assumed to remain constant (i.e.
the relative reflection $R$ is variable). 

2) a change of the intrinsic slope as well as the reflected component
(solid line). In this case the reflection component is assumed to promptly
respond to the intensity variation of the continuum (i.e. relative reflection 
$R$ is constant).

Looking at Figure \ref{spec ratios}, it is clear that the limited
statistics does not allow us to distinguish between the two cases of 
constant or variable reflection.
The observed behaviour can be reproduced by 
a variation of the intrinsic slope.
This will be confirmed with a spectral analysis on the separate flux 
states, discussed in the next section.

\subsection{Spectral variability: the different flux level spectra}
\label{spec var2}

We performed a spectral analysis of 
the different flux spectra,
extracted as shown in Figure \ref{light} and in 
Table \ref{journal}.
The best fit model employed in this analysis is the one giving the best
fit to the integrated spectra, i.e. the model F. 

\begin{figure*}
\centering
\includegraphics[height=11.0cm,width=8.5cm,angle=-90]{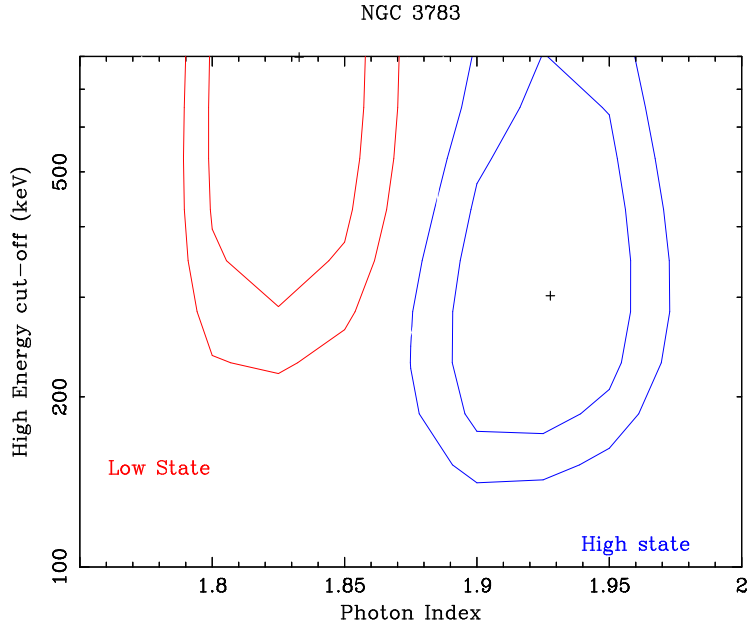}
\caption[]{NGC 3783. 1$\sigma$ and 2$\sigma$ confidence levels E$_c$ vs 
$\Gamma$ in the high (H) and low (L) states.}
\label{ec_gamma}
\end{figure*}

We confirm that the observed variations can be explained by a change of the 
intrinsic continuum. In particular the source steepens 
($\Delta\Gamma\sim0.1$, see Figure \ref{ec_gamma}, Table \ref{fit var}) 
when brightens. 
The limited statistics in the separate flux states did not allow us 
to detect variations of E$_c$ (see contour plot in Figure \ref{ec_gamma}).
No variations in the ionization status of the warm gas was observed.

%%%%%%%%%%%%% TABLE BEST FIT MODEL F IN THE DIFFERENT STATES %%%%%%%

\begin{table*}
\caption{NGC 3783: Spectral variability, best fit parameters in the 
different flux states in the model F.} 
\begin{flushleft}
\begin{tabular}{||c|c|c|c||} \hline
~ & ~ & & \\

~ & & \bf Low State: L & \bf High State: H  \\

~ & ~ & & \\ \hline

~ & ~ & & \\

{\bf Warm} & $\lg$(U) & 1.28$\pm^{0.13}_{0.07}$ & 1.38$\pm^{0.15}_{0.19}$ \\ 
\cline{2-4}

~  & ~ & & \\ 

{\bf Absorber} & $\lg (N_H)$ & 22.27\p0.11 & 22.32$\pm^{0.12}_{0.10}$\\
\hline

~  & ~ & & \\

{\bf Intrinsic} & E$_c$ (keV) & $>$ 250 & 300$\pm^{500}_{150}$
\\ \cline{2-4}

~ & & & \\
{\bf Continuum} & $\Gamma$ & 1.83\p0.04 & 1.93$\pm^{0.04}_{0.06}$ \\ 
\cline{2-4}

~  & ~ & & \\

~ & $^\diamond$Norm. & 1.9 & 2.5 \\ \hline
~  & ~ & &\\

{\bf Reflection} & $^\circ$R & 0.58$\pm^{0.29}_{0.34}$ & 0.77$\pm^{0.47}_{0.23}$\\ 
\cline{2-4}

~ & ~ & & \\

{\bf and} & E$_L$ (keV) & 6.39$\pm$0.07 & 6.39\p0.06 \\ \cline{2-4}

~  & ~ & & \\

{\bf Iron Line} & $\sigma_L$ (keV)& 0.37$\pm^{0.20}_{0.17}$ & 
0.25$\pm^{0.24}_{0.22}$ \\ \cline{2-4}
~  & ~ & &\\

~& EW$_L$ (eV)& 204\p78 & 163\p70\\ \hline
~  & ~ & & \\

{\bf OVII} & E$_2$ (keV) & 0.5\p0.2 & 0.5\p0.2\\ \cline{2-4}
 ~ &  ~ & &\\

{\bf Line} & $\sigma_2$ (keV) & 0 (frozen)  & 0 (frozen)\\ \cline{2-4}
~ & ~ &  & \\

~ & EW$_O$ (eV) & 34\p51 & 30\p41 \\ \hline
 ~ &~ & &\\
{\bf Soft} & kT (keV) & 0.21$\pm^{0.02}_{0.03}$ & 0.22$\pm^{0.11}_{0.13}$ \\ 
\cline{2-4}
 ~ &~ & & \\
{\bf Emission} & $^\star$Norm. & 4.5$\pm^{1.2}_{1.1}$ & 4.6$\pm^{2.2}_{0.9}$
 \\ \hline
 ~ &~ & & \\
& $\chi^2$/dof & 85.26/102 & 73.20/102 \\ \hline
\noalign{\hrule}
\noalign{\medskip}
\end{tabular}
\newline
\small{$^\diamond$ In 10$^{-2}$ photons keV$^{-1}$ cm$^{-2}$ s$^{-1}$,
a 1 keV.}
\newline
\small{$^\circ$ with cos(i)=30$^\circ$} 
\newline
\small{$^\star$ In 10$^{-4}L_{39}/D_{10}^2$, where L$_{39}$ is 
the source luminosity in units of $10^{39}$ ergs/sec and D$_{10}$ is 
the distance to the source in units of 10 kpc.}
\end{flushleft}
\label{fit var}
\end{table*}

\section{Discussion}
\label{discussion}

\subsection{Intrinsic Continuum and spectral variability}
\label{ic}

The presence of a high energy cut-off 
was required to fit the X-ray/$\gamma$-ray spectra of a sample of 
Seyfert observed with different satellites (\cite{ZZ95}).
The $E_C=340\pm^{560}_{109}$ keV detected in \37 by BeppoSAX is consistent
with the values observed in other Sy 1s of the BeppoSAX sample 
(ranging between 70 keV up to 300 keV, Perola \etal in preparation).

The spectral variability detected in our observation can be accounted
for by variations of the intrinsic slope. In particular the intrinsic spectrum
steepens when the source brightens ($\Delta \Gamma\sim 0.1$ with 
$(\Delta F/F)_{2-10 keV} \sim 20\%$).
This behaviour was observed also in other sources: NGC 5548 (\cite{N00}), 
NGC 4151 (Piro 2002), NGC 7469 (Nandra \etal 2000), MCG-6-30-15 (Vaughan \& 
Edelson 2001), and IC 4329A (Done \etal 2000).

A two-phase model involving a hot corona emitting medium-hard 
X-rays by Comptonization and a cold optically thick layer which provides 
the soft photons to be Comptonized (Haardt \& Maraschi 1991, 
Haardt \etal 1997), 
can explain the observed intrinsic variability.
In this model the physical parameters of the corona can drive the 
spectral variations observed in the hard-medium energy range, while 
those in the soft band are driven by the temperature $T_{BB}$ of
the soft photons emitted by the cold layer.

If the corona is not pair dominated, i.e. the compactness 
parameter $l_c=\frac{L_c \sigma_T}{c\Delta t (mc^3)} \le 10$, 
the two phases  model predicts a noticeable spectral variability
even when the luminosity variation is less than a factor two.
Our variability study suggests that this is the case for \37. 
We measure $\Delta\Gamma\sim 0.1$ (see Figure \ref{ec_gamma} and 
Table \ref{fit var}), with no variation of the Comptonization luminosity 
(0.1-200 keV) in the H and L states: 
$L_c^H \sim L_c^{L}\sim 10^{44}$ erg/s.
We can put a lower limit to the compactness parameter. 
The time scale over which the flux changes by a factor 2 is
greater than 5 days, so we obtain $l_{c} \ge 0.2$.
A detailed Comptonization code was employed to fit the 
data of the BeppoSAX observation of NGC 5548 (\cite{pop2000}). 
The spectral variability observed in that source could by accounted for by 
a change of the spectral slope $\Delta \Gamma\sim 0.2$ with 
small hard luminosity changes. 
The spectral softening in the high state is explained (when a realistic 
model of Comptonization is employed) by a decrease 
of kT$_e$ due to an increase of the soft photon flux.
The anticorrelation between the slope and the high energy cut-off is not
revealed when a cut-off power law is used as a zero order approximation of 
a Comptonized spectrum.
Also in the case of NGC~5548 the trend of variability suggests a hot plasma
which is not pair-dominated (and the estimated lower limit for the 
compactness was $l_c \ge 0.2$). 
In the case of \37 an accurate Comptonization model was employed only to 
fit the total spectrum, with no spectral variability analysis (\cite{pop2001}).
The estimated coronal parameters are substantially different between 
the two models (the coronal temperature kT$_e=265\pm^{15}_{5}$ keV 
is higher than that measured with our simpe cut-off power law model
 E$_c$/2). 
This is due to the presence of an energy break in the 
Comptonization model, produced when the anisotropy of the 
soft photon field is taken into account.
   
\subsection{Warm absorber and soft emission}
\label{se}

The low energy spectrum of \37 is characterized by the presence of
the warm absorber. The complex structure of this gas was investigated
in the Chandra observation (\cite{kaspi2}).
The model consisted of two absorption/emission components with the same column 
density $\lg(N_H)=22.1$, but different ionization parameter:
$\lg(U_{oxygen})= -0.745$, with a covering 
factor f=0.3 and $\lg(U_{oxygen})= -1.745$, with a covering 
factor f=0.5 (these values are computed between 0.538 keV and 10 keV).
We fitted the BeppoSAX data with a grid of single-zone warm absorber models
build with CLOUDY (\cite{ferland97}).
These models include only the transmitted spectra, and do not include 
the contributions of both gas emission and resonant absorption.
The ionization parameter, computed from $E>$ 1 Ryd, in the total spectrum
was $\lg(U)= 1.33$ (see Table \ref{fit tot2}). 
Taking into account the ionizing continuum used to fit Chandra 
data, we get the relation U$_{oxygen}$=0.114 U(0.1-10 keV)
(\cite{kaspi2}); in the case of BeppoSAX spectrum we employed 
the ionizing continuum from Mathews \& Ferland (1995) 
(listed in Table \ref{MF_continuum}) to rescale the 
value of U($>$ 1 Ryd) in the 0.1-10 keV energy range. In this case we 
found  U($>$ 1 Ryd) $\sim$ 50 U(0.1-10 keV) (see also George \etal 1998a).
The ionization parameter during our observation is marginally consistent with 
that of the low ionization gas observed in the Chandra spectrum 
$U_{SAX}$(0.1-10 keV)=0.4$\pm$0.1. 
No variation of the low-ionization component was observed between 
BeppoSAX and Chandra observations, not surprisingly as no flux changes 
in 0.5-2 keV (F$_{0.5-2 keV}=2\times 10^{-11} \flux$) between the two 
observations have been detected.

The high ionized absorption features of the heavier elements produced in the 
more ionized component are too small to be observed given the limited LECS 
sensitivity and energy resolution.
The warm absorber parameters obtained with BeppoSAX are derived mainly
from the deep OVII and OVIII absorption edges produced in 
the low ionization gas.
The OVII triplet observed by Chandra at $\sim 0.5-0.6$ keV is only 
marginally required by our data (P$_F$=77\%). 
This feature was also observed in the BeppoSAX 
spectrum of NGC 5548 (\cite{N00}), and later on confirmed by Chandra
(Kaastra \etal 2000).
An additional absorption line (at $\sim$ 1 keV) 
corresponding to the blend of many iron L-shell absorption lines 
observed by Chandra, is very marginally required by the model (P$_F$=52\%).
When we add this component the warm absorber parameters as well as 
those of the intrinsic continuum do not change.

A reanalysis of the ASCA observation of 1996 taking into account (1) the
inclusion of the reflection component as measured by the RXTE observation 
simultaneous with those Chandra in 2000 and (2) the use of the Yaqoob (2000) 
method to account for the degradation of the SIS detectors, was
performed by Kaspi \etal (2001). The intrinsic continuum they found
is fully consistent with that measured by BeppoSAX
($\Gamma$=1.85\p0.03, A(1 keV)
=2.21$\times 10^{-2}$ ph cm$^{-2}$ s$^{-1}$ keV$^{-1}$), and the 
warm absorber is consistent with the low ionization component observed by 
Chandra ($U_X$=0.133\p0.003, $\lg N_H$=22.15\p0.03). 

A significant soft emission well modeled by a black body component with 
a temperature kT=220$\pm$30 eV was detected in the BeppoSAX observation.
No variation of the soft component emission was detected.
We may associate this emission to the accretion disc around 
the central source.
The IUE monitoring of \37 from 1978 to 1985 detected a UV-bump
from 5150 $\AA$ up to the soft X-rays. 
The observed bump could be attributed to a thermal emission from an 
optically thick medium.
When the UV bump was reproduced with a sum of blackbody spectra, the
highest disc temperature was 280,000 K (\cite{alloin95}).

\begin{table}[h]
\caption{Ionized continuum from Mathews \& Ferland (1987) employed 
to build a grid of single-zone warm absorber model to fit the low energy 
spectrum of \37} 
\begin{flushleft}
\begin{tabular}{||c|c|c||} 
\hline
$\nu$(Ryd) & $\lg(F_\nu)$ & slope $\alpha$ \\ 
\hline
0.206 & 2.6576 & 0.5 \\
1.743 & 2.194 & 1 \\
4.130 & 1.819 & 3 \\
26.84 & -0.6192 & 0.7 \\
\noalign{\hrule}
\noalign{\medskip}
\end{tabular}
\end{flushleft}
\label{MF_continuum}
\end{table}

\subsection{Compton reflection and Iron line}
\label{refl disc}

\begin{figure*}
\centering
\includegraphics[height=11.0cm, width=8.5cm,angle=-90] {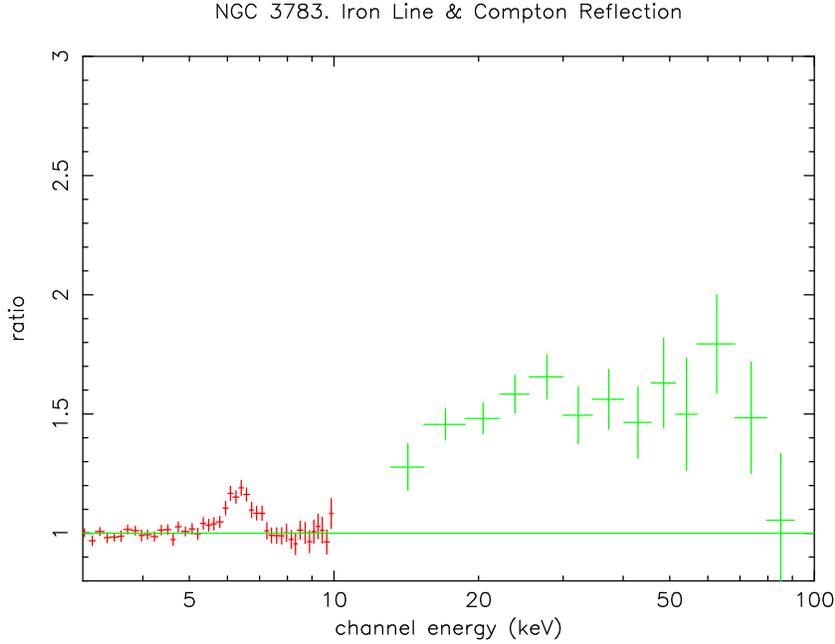}
\caption[]{Ratio data/best fit model when the normalization of the
iron line and the Compton reflection component are set equal zero.}
\label{refl1}
\end{figure*} 

\begin{figure*}
\centering
\includegraphics[height=8.0cm, width=6.0cm,angle=-90] {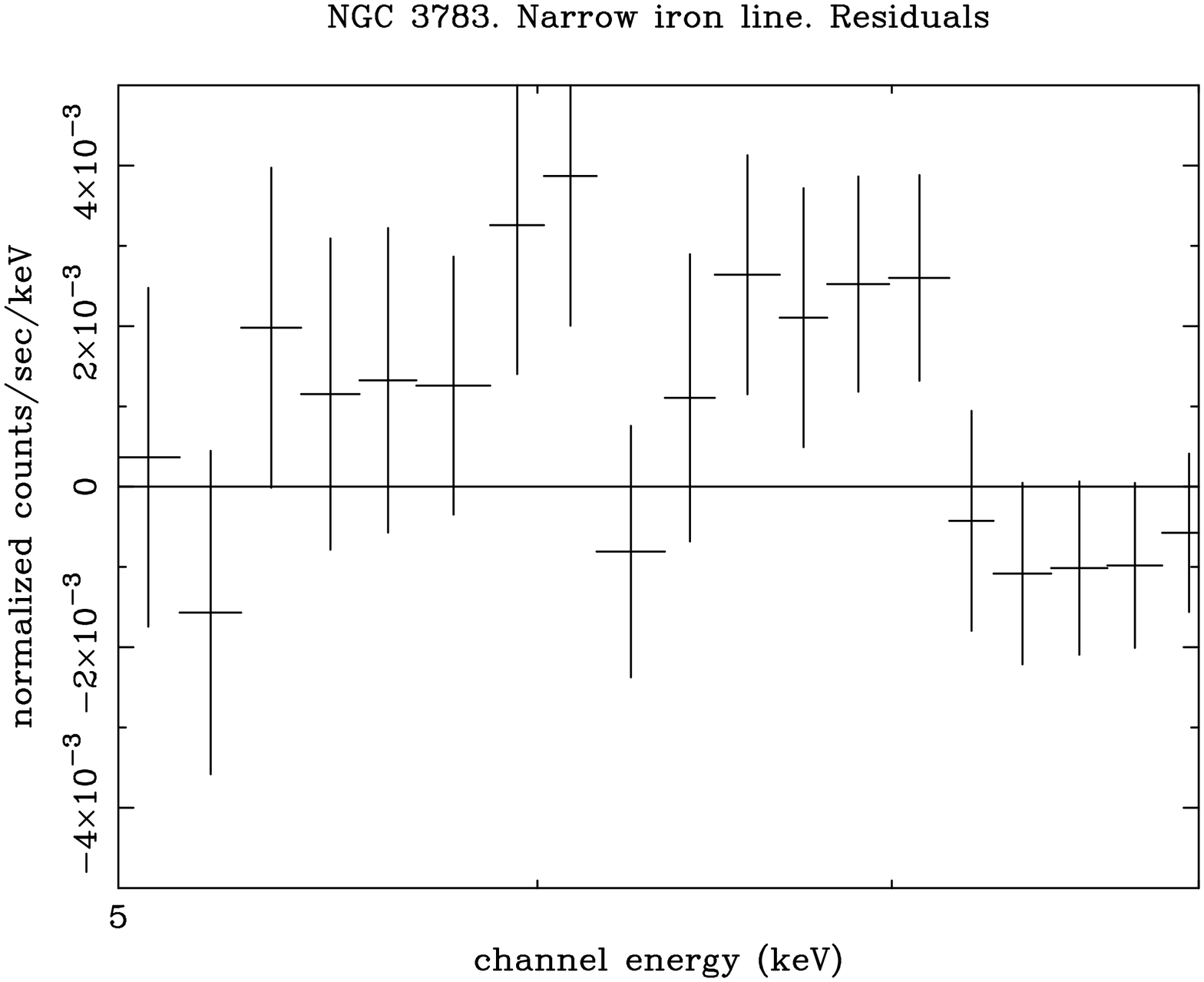}
\includegraphics[height=8.0cm, width=6.0cm,angle=-90] {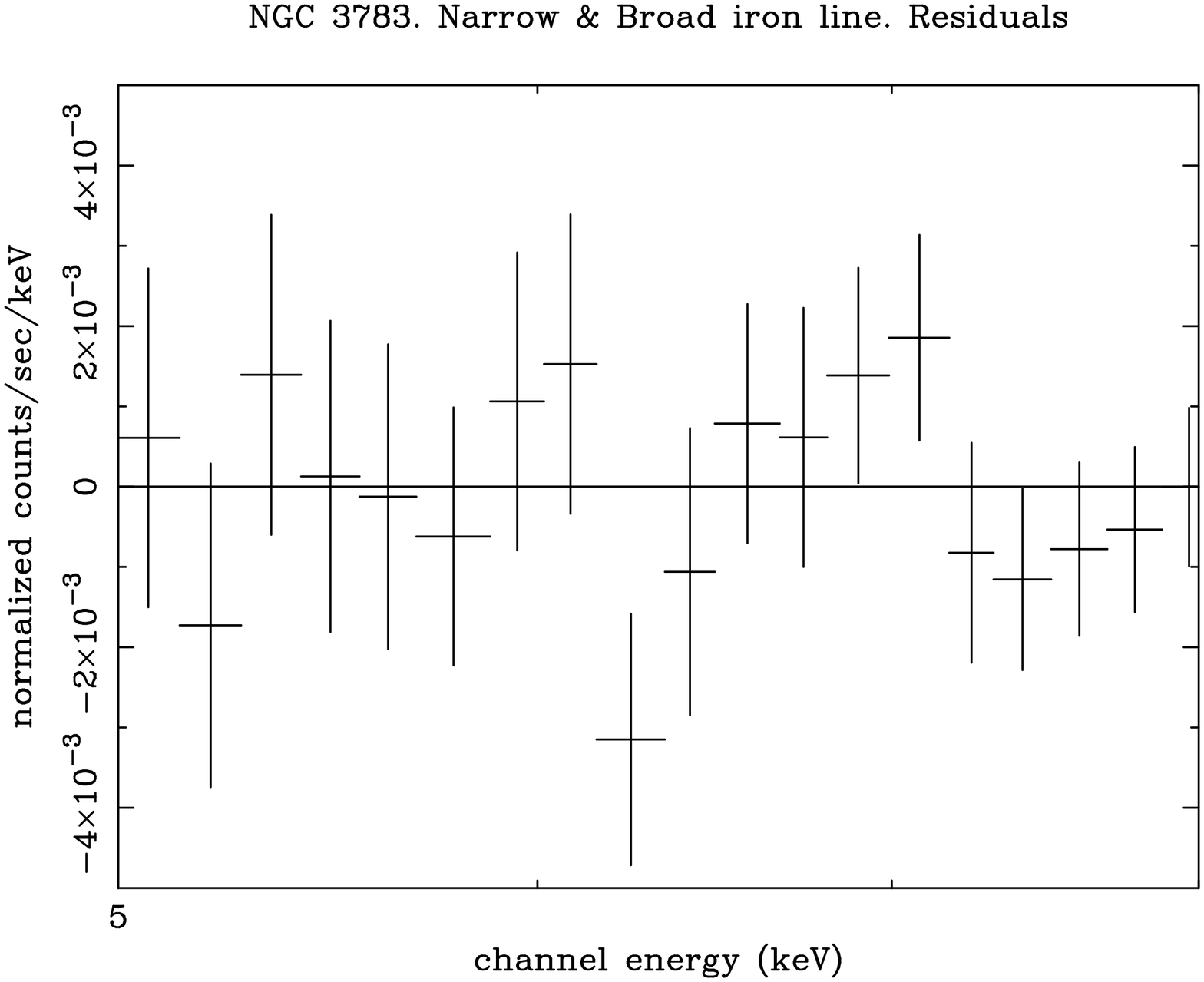}
\caption[]{Left panel. Residuals to narrow iron line when all the line
parameters fixed to those of Chandra observation.
Right panel. A similar plot to left panel with the addition 
of a broad gaussian component.}
\label{iron1}
\end{figure*} 

\begin{figure*}
\centering
\includegraphics[height=11.0cm, width=8.5cm,angle=-90] {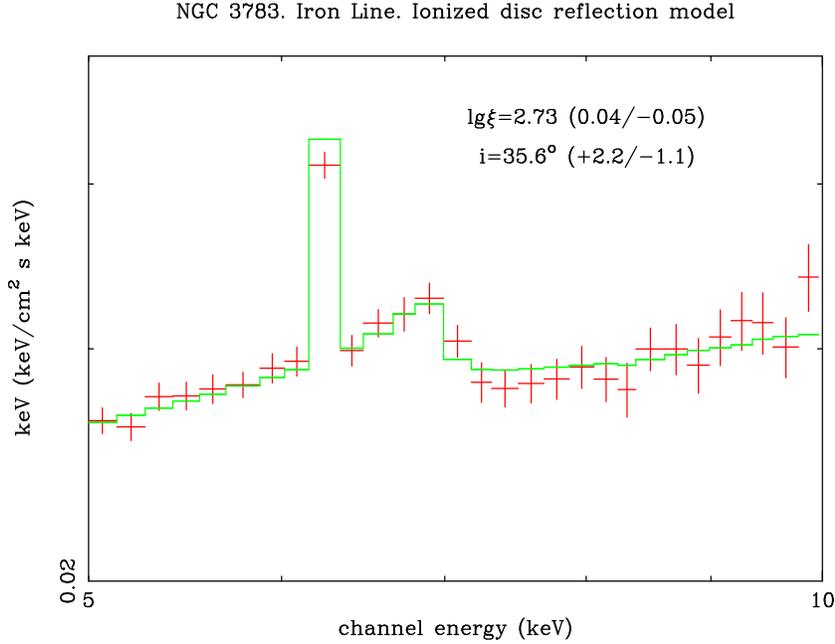}
\caption[]{The unfolded BeppoSAX/MECS spectra (EF$_E$) of \37 is plotted
when the ionized disc reflection model of Ross \& Fabian (1993) is employed 
to fit the data.
A narrow iron line component with all the parameters linked to those 
observed by Chandra is added to the model. A broad disc component is 
also clearly present.}
\label{iondisc}
\end{figure*} 

The broad band capability of BeppoSAX allowed us to detect the 
Compton reflection together with a broad Fe K$\alpha$ emission line
(Figure \ref{refl1}). 
The energy of the iron line $E_{Fe}=6.39 \pm 0.09$ keV, 
is consistent with FeI-XII.
In the model F (see Table \ref{fit tot2}) this features is well reproduced by a broad gaussian 
component with $\sigma_{Fe}=0.36\pm^{0.09}_{0.08}$ keV and 
EW$_{Fe}$=190\p47 eV.
A broad and asymmetric line is expected if the X-ray reflection
is occurring in the inner region of an accretion disc (Fabian \etal 2000, 
Matt \etal 1992).
However there is evidence that in NGC 3783 
as well as in other Seyfert 1 galaxies, the line is actually composed 
of two components, one broad and variable and the other narrow and constant.
To support this idea we recall that in the 
Chandra observation on \37
a narrow unresolved iron line (with EW=115\p36 eV) was detected 
(Kaspi et al. 2001). 
The upper limit for the FWHM was 3250 km s$^{-1}$, consistent with 
this line originating outside the BLR. The line flux is consistent with 
 predictions from models of torus emission. 
The Chandra data could not constrain any broad Fe K$\alpha$ emission 
though they were consistent with the broad line found by ASCA in this source
(EW=230$\pm$70 eV, \cite{G98b}).

If we fit the iron line with a narrow component with the parameters fixed to
those detected by Chandra (E$_{nl}=6.39$ keV, 
I$_{nl}=6.6\times10^{-5}$ ph cm$^{-2}$ s$^{-1}$), we find strong residuals 
around 6 and 7 keV (see Figure \ref{iron1}, left panel). 
If we add a second broad gaussian component the residuals disappear 
(see right panel in Figure \ref{iron1}). 
The broad component is required at 99.9\% c.l. 
(following a F-test adding three interesting parameters), 
with  $\sigma_{bl}=0.72\pm^{1.28}_{0.27}$ keV EW$_{bl}$=115 \p 76 and 
E$_{bl}=6.27\pm^{0.25}_{0.30}$ keV, i.e. corresponding to neutral iron.
These values are fully consistent with those of the ASCA observation
in 1996 when the ASCA data were modeled assuming the underlying continuum 
derived by the analysis of a RXTE observation (simultaneous with those of
Chandra in 2000, \cite{kaspi2}), to constrain the reflection component.

The simultaneous BeppoSAX/XMM observation of the NLSy1 NGC~5506 
(\cite{matt01b}) showed that in that case the bulk 
of the Compton reflection has to be associated with the narrow iron 
line, as in  the the case of NGC~4051 (Guianazzi \etal 1998).
In other objects observed by XMM 
(Mkn~509, Pounds \etal 2001; Mkn~205, Reeves \etal 2001), evidence for 
ionized accretion disc was also found.
An ionized disc reflection model was also employed to fit the data of \37.
This source was included in a small sample of Sy 1s observed by 
BeppoSAX analysed with the ionized disc reflection model of Ross \& Fabian 
(1993) (\cite{me_estec}).
In this model the most important quantity in determining 
the shape of the reflected continuum is the ionization parameter 
$\xi=4\pi F_x/n_H$, where F$_x$ is the X-ray flux (between 0.01-100 keV) 
illuminating a slab of gas with solar abundances and constant hydrogen 
number density n$_H$=10$^{15}$ cm$^{-3}$.
The K$\alpha$ iron emission line as well as the emission features 
from the disc at low energy, are included in the model.
In the fit, we applied to the spectrum a 
relativistic blurring appropriate for a Schwarszchild geometry assuming a 
disc emissivity law (Fabian \etal 1989). 
During the fit a narrow iron line component with all the 
parameters linked to those of Chandra observation was included.
This analysis showed that the reflected spectrum of \37 can be produced 
($\chi^2$/dof=104/101) by a highly ionized disc characterized by:
inner radius (pegged to the lower value) r$_{in}$=6 r$_g$, outer radius 
r$_{out}$= 7$\pm^{1}_{0.3}$ r$_g$ with r$_g$=GM/c$^2$ the gravitational radius,
ionization parameter $\lg\xi=2.73\pm^{0.04}_{0.05}$ and inclination angle 
$i=35.6\pm^{1.1}_{2.2}$ degrees (see Figure \ref{iondisc}).
In this case the observed soft X-rays excess can be accounted for by the disc 
emissivity and no additional component at E $<$ 2 keV is required.
This suggests that the Compton reflection observed in \37 can be completely 
associated with the broad Fe component supposing that both were created in 
a relativistic disc. Unfortunately,
the energy resolution and sensitivity of BeppoSAX are not
adequate to disentangle both the Fe line components and simultaneously the
fraction of reflection associated with them.
A simultaneous BeppoSAX/XMM observation would be the best way to achieve this 
purpose.
 
\section {Summary}
\label{summary}

We have analysed the long BeppoSAX observation of the Seyfert 1 galaxy 
NGC~3783.
We detect a high energy cut-off of the intrinsic power; 
the value of E$_c$=340$\pm^{560}_{107}$ keV is consistent with those 
observed so far in a sample of Seyfert galaxies, which ranges from 
70 keV up to 300 keV. Spectral variability was clearly present, well
modeled by a steepening of the 
intrinsic continuum ($\Delta \Gamma \sim 0.1$) with no total luminosity 
change. This behaviour is consistent with 
a two-phase disc-corona model in which with the hot plasma is not 
pairs dominated (as observed also in the Seyfert 1 galaxy NGC~5548). 

The presence of a warm gas is confirmed by our analysis. 
Its ionization status is consistent with the low-ionization component 
detected by Chandra. 
NGC~3783 is one of the few sources in which BeppoSAX detected a soft excess.
We reproduced the soft emission with a black body component with a temperature
kT$\sim$0.2 keV.

The Compton reflection hump and the iron emission line are clearly 
detected.
In addition to the narrow unresolved iron component observed by Chandra,
a broad iron line component ($\sigma=0.72\pm^{1.28}_{0.27}$ keV) is required 
to fit the BeppoSAX data.
The observed reprocessing features can be reproduced with a highly ionized 
accretion disc ($\lg\xi=2.73\pm^{0.04}_{0.05}$). In this case 
the soft excess observed at E $<$ 2 keV can be accounted for by the disc 
emissivity and no additional soft X-rays component is required to 
fit the spectrum.

\begin{acknowledgements}
We gratefully acknowledge A.C. Fabian and D.R. Ballantyne for 
providing us with the ionized disc reflection code. 
A.D.R. would like to thank G.C. Perola for useful discussions. 
G.M. and F.F. acknowledge financial support from ASI and MURST (under
grant {\sc cofin-00-02-36}). 
We thank the SAX Scientific Data Center and
the anonymous referee who provided helpful suggestions and comments.

\end{acknowledgements}

\end{document}